\documentclass[apj]{emulateapj}

\shorttitle{Laboratory photo-chemistry of pyrene clusters}
\shortauthors{Zhen et al.}

\begin{document}
\title{Laboratory photo-chemistry of pyrene clusters: an efficient way to form large PAHs}
\author{Junfeng Zhen$^{1,2,*}$, Tao Chen$^{3,4}$, Alexander G. G. M. Tielens$^{3}$} 

\affil{$^{1}$CAS Key Laboratory for Research in Galaxies and Cosmology, Department of Astronomy, University of Science and Technology of China, Hefei 230026, China} 
\affil{$^{2}$School of Astronomy and Space Science, University of Science and Technology of China, Hefei 230026, China}
\affil{$^{3}$Leiden Observatory, Leiden University, P.O.\ Box 9513, 2300 RA Leiden, The Netherlands}   
\affil{$^{4}$School of Engineering Sciences in Chemistry, Biotechnology and Health, Department of Theoretical Chemistry \&\ Biology, Royal Institute of Technology, 10691, Stockholm, Sweden}
\email{jfzhen@ustc.edu.cn}

\begin{abstract}
In this work, we study the photodissociation processes of small PAH clusters (e.g., pyrene clusters). The experiments are carried out using a quadrupole ion trap in combination with time-of-flight (QIT-TOF) mass spectrometry. The results show that pyrene clusters are converted into larger PAHs under the influence of a strong radiation field. Specifically, pyrene dimer cations (e.g., [C$_{16}$H$_{10}$$-$C$_{16}$H$_{9}$]$^+$ or C$_{32}$H$_{19}$$^+$), will photo-dehydrogenate and photo-isomerize to fully aromatic cations (PAHs) (e.g., C$_{32}$H$_{16}$$^+$) with laser irradiation. The structure of new formed PAHs and the dissociation energy for these reaction pathways are investigated with quantum chemical calculations. These studies provide a novel efficient evolution routes for the formation of large PAHs in the interstellar medium (ISM) in a bottom-up process that will counteract the top-down conversion of large PAHs into rings and chains, and provide a reservoir of large PAHs that can be converted into C$_{60}$ and other fullerenes and large carbon cages.
\end{abstract}


\section{Introduction}
\label{sec:intro}
Mid-infrared (IR) spectra of the interstellar medium (ISM) are dominated by broad features at 3.3, 6.2, 7.7, 8.6 and 11.2 $\mu$m that are generally attributed to IR fluorescence from UV-pumped Polycyclic Aromatic Hydrocarbon molecules (PAH) \citep{sel84,pug89,all89}. In addition, the IR signatures of the fullerene, C$_{60}$, have also been identified at 7.2, 8.5, 17.4 and 19.0 $\mu$m \citep{cam10,sel10}. Interstellar IR spectra also show evidence for the presence of PAH clusters and very small dust grains \citep{rapacioli05a,berne07}. PAHs and PAH clusters are ubiquitous and abundant, containing some 10 \% of the elemental carbon in space. PAHs are thought to play an important role in the ionization and energy balance of the interstellar medium of the galaxy \citep[and references therein]{tie08}. Furthermore, the IR emission bands are also prominent in protoplanetary disks around young stars and thus an important component of the organic inventory of regions of planet formation \citep{taha2018,doucet2007}. Finally, PAH molecules, as well as C$_{60}$, are known to be an important component of meteorites in the solar system \citep{bec94,sep08}.

PAHs and related species are thought to undergo a complex chemical evolution in the harsh radiation field of PDR \citep{berne12,rapacioli06}. In competition with ionization and IR fluorescence, highly excited PAHs may fragment \citep{lea86,lep01,zhen2016}. Under intense UV radiation, experiments show that this will lead to a complete stripping of the H's from a PAH followed by C-loss, ring \&\ chain formation and conversion of some of the pure carbon flakes to cages and fullerenes \citep{zhen2014a,zhen2014b}. As an intermediate step in this process, fragmentation and isomerization processes may convert PAHs and PAH clusters to large stable species; the so-called grandPAH hypothesis \citep{tie13, and2015, zhen2018}. In this top-down chemical route towards molecular complexity in space, PAHs injected into the ISM by Asymptotic Giant Branch stars as part of the soot formation process \citep{frenklach89,cherchneff92} are broken down to smaller and smaller species \citep{berne12,tie13,zhen2014b}. Formation of PAHs (and fullerenes) from small hydrocarbons, in a bottom-up fashion, is highly inefficient in the ISM \citep{hen86,bet1997}. However, the UV-driven breakdown of PAHs may be counteracted in the ISM through energetic processing of PAH clusters. Specifically, laboratory studies of processing of van der Waals clusters of PAHs and fullerenes by energetic ions have revealed that direct knock-out of carbon atoms leads to the formation of chemically bonded large species \citep{del15,chen2015,zet13}

PAH clusters or related complex species (e.g., ion-neutral dimers) are believed to be the self-assembled intermediaries between free gas-phase PAHs and amorphous carbon particles, and have been considered as an important role of the interstellar PAHs model \citep{all89,rapacioli05a,rapacioli06,rhee07}. In the intense and high energy radiation field of photo-dissociation regions (PDRs), the concentration of PAH cations will be relatively high \citep{all85,all89}. Significant concentrations of ion-neutral dimers are likely to be present as well, in steady-state with the monomers, similar to the PAH neutral and ion balance \citep{all89, bak98, li01}. The neutral-ion dimer concentration will strongly depend on the flux of ionizing radiation \citep{dar00,wit06,rapacioli06}. Quantum chemical studies show that various PAH molecules can form cationic dimers with binding energies intermediate between normal van der Waals complexes and chemical bonds \citep{rhee07,dot16}. Interestingly, it is speculated that the fluorescence from PAH aggregates is the main cause of the extended red emission (a broad luminescence process in the ISM), and it is also the cause of widespread emission in sooting flames \citep{mil05,rhee07}.

It is clear that interstellar PAHs and PAH-related species are expected to be greatly influenced by strong radiation fields in space \citep{lea86,ver90,bak94,lep01}). However, past experimental studies have focused on the top-down process and the issue of reformation of large PAHs under relevant interstellar conditions (i.e., the bottom-up process) has not received much attention. As these two processes will be in competition in space, both aspects must be considered at the same time. In this study, we investigate the photochemical processing of pyrene cluster cations. This paper is organized as follows: The experimental techniques are described in section~\ref{sec:exp}, the experimental and theoretical results are analyzed and discussed in section~\ref{sec:results} and \ref{sec:theoretical}, and the astronomical implications of the results are illustrated in section~\ref{sec:discussion}. In the end, a brief conclusion is provided in section~\ref{sec:concl}

\begin{figure*}[t]
 \centering
 \includegraphics[width=\textwidth]{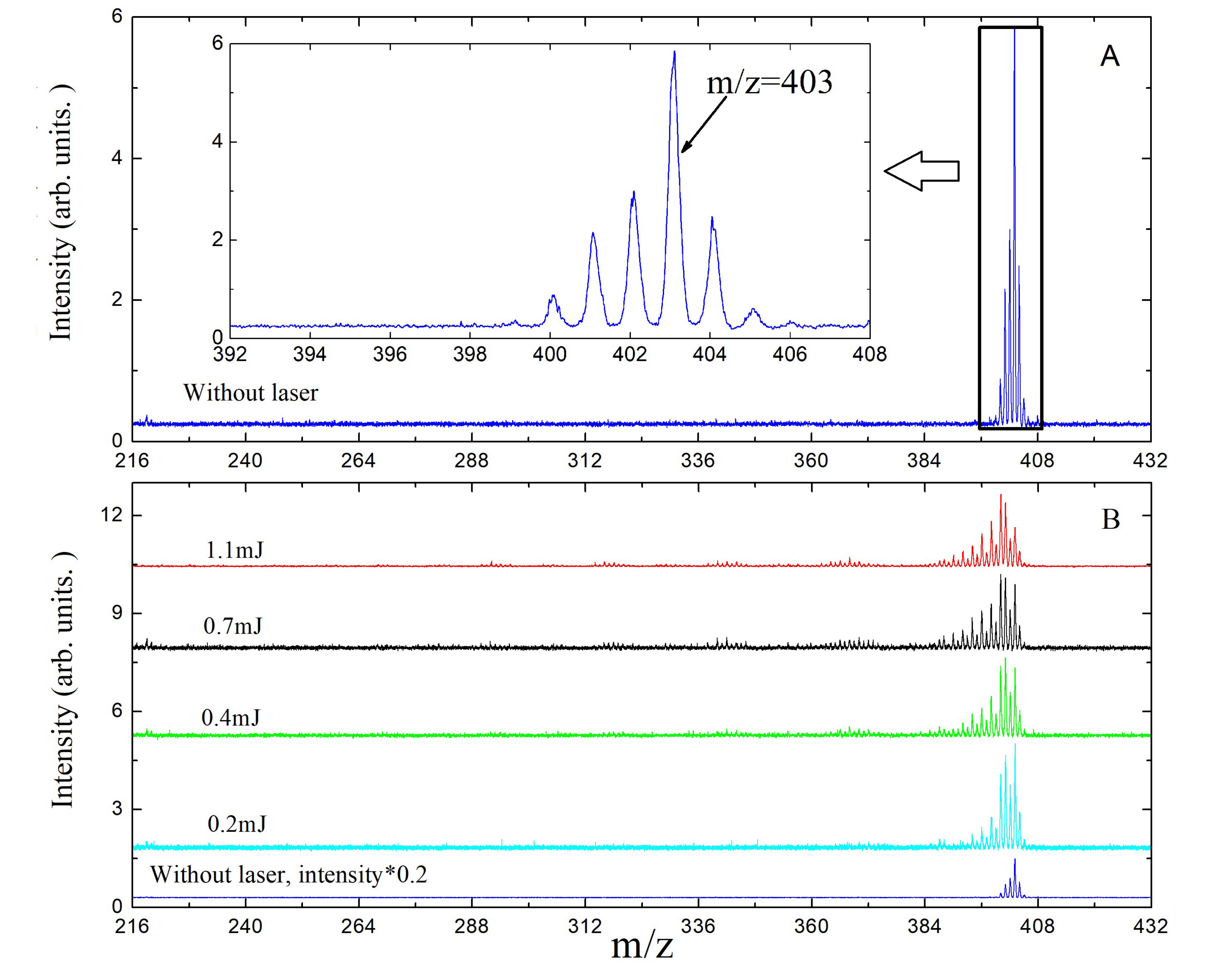}
 \caption{Upper panel (A): Mass spectrum of pyrene dimer complex cations trapped in QIT after SWIFT isolation and before laser irradiation. The inset is a zoom-in mass spectrum, revealing the presence of a number of pyrene dimer complexes. Lower panel (B): Mass spectrum of pyrene dimer complex cations before and after irradiation at 595 nm with different laser energies. 
 }
 \label{fig1}
\end{figure*}

\begin{figure*}[t]
 \centering
 \includegraphics[width=\textwidth]{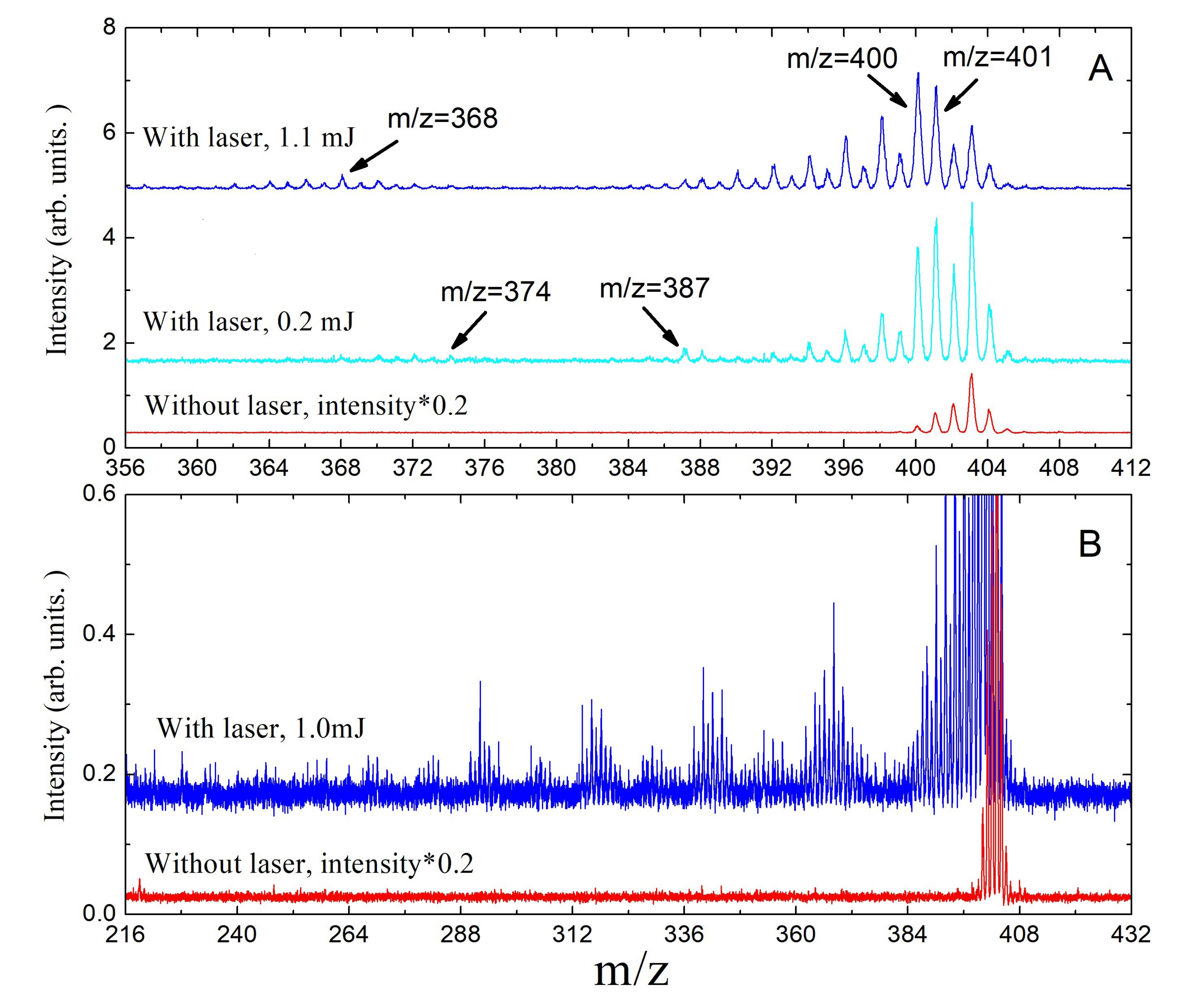}
 \caption{Mass spectrum of pyrene dimer complex cations: the upper panel (A) is without and with 0.2 and 1.0 mJ laser irradiation in the range of m/z=356-412. The lower panel (B) is before and after 1.0 mJ laser irradiation in the range of m/z=216-432.
 }
 \label{fig2}
\end{figure*}

\begin{figure*}[t]
 \centering
 \includegraphics[width=\textwidth]{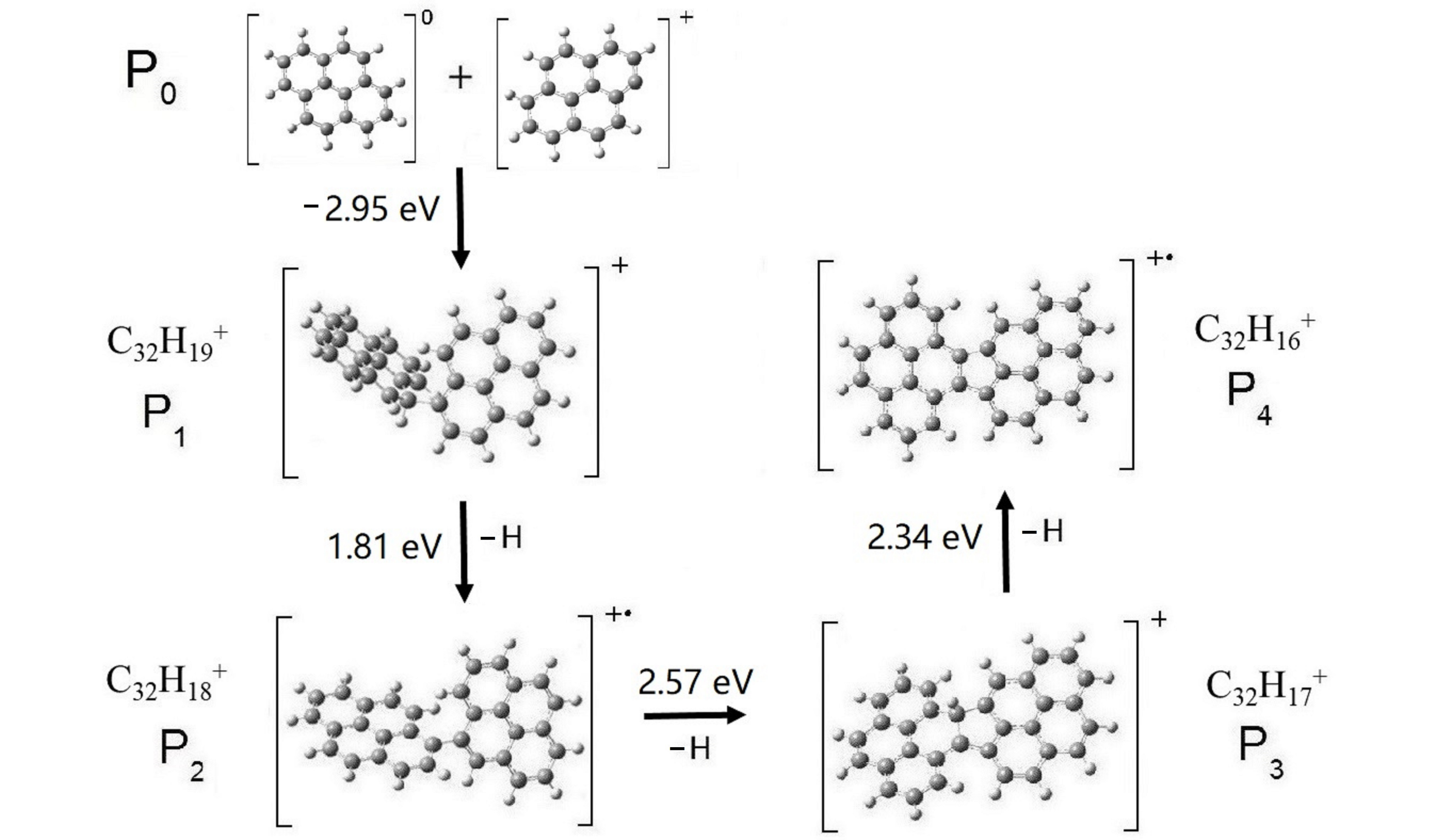}
 \caption{Theoretical reaction pathway and fragmentation map of pyrene dimer complex cations (C$_{32}$H$_{19}$$^+$, P$_1$). From left to right, the optimized structures from P$_{0}$, to P$_{1}$ (C$_{32}$H$_{19}$$^+$), to P$_2$ (C$_{32}$H$_{18}$$^+$), to P$_3$ (C$_{32}$H$_{17}$$^+$), to P$_4$ (C$_{32}$H$_{16}$$^+$). The arrows are labeled by the dissociation energies of the dimer or the H-losses in these the reactions.
 }
 \label{fig3}
\end{figure*}

\begin{figure}[t]
 \centering
 \includegraphics[width=\columnwidth]{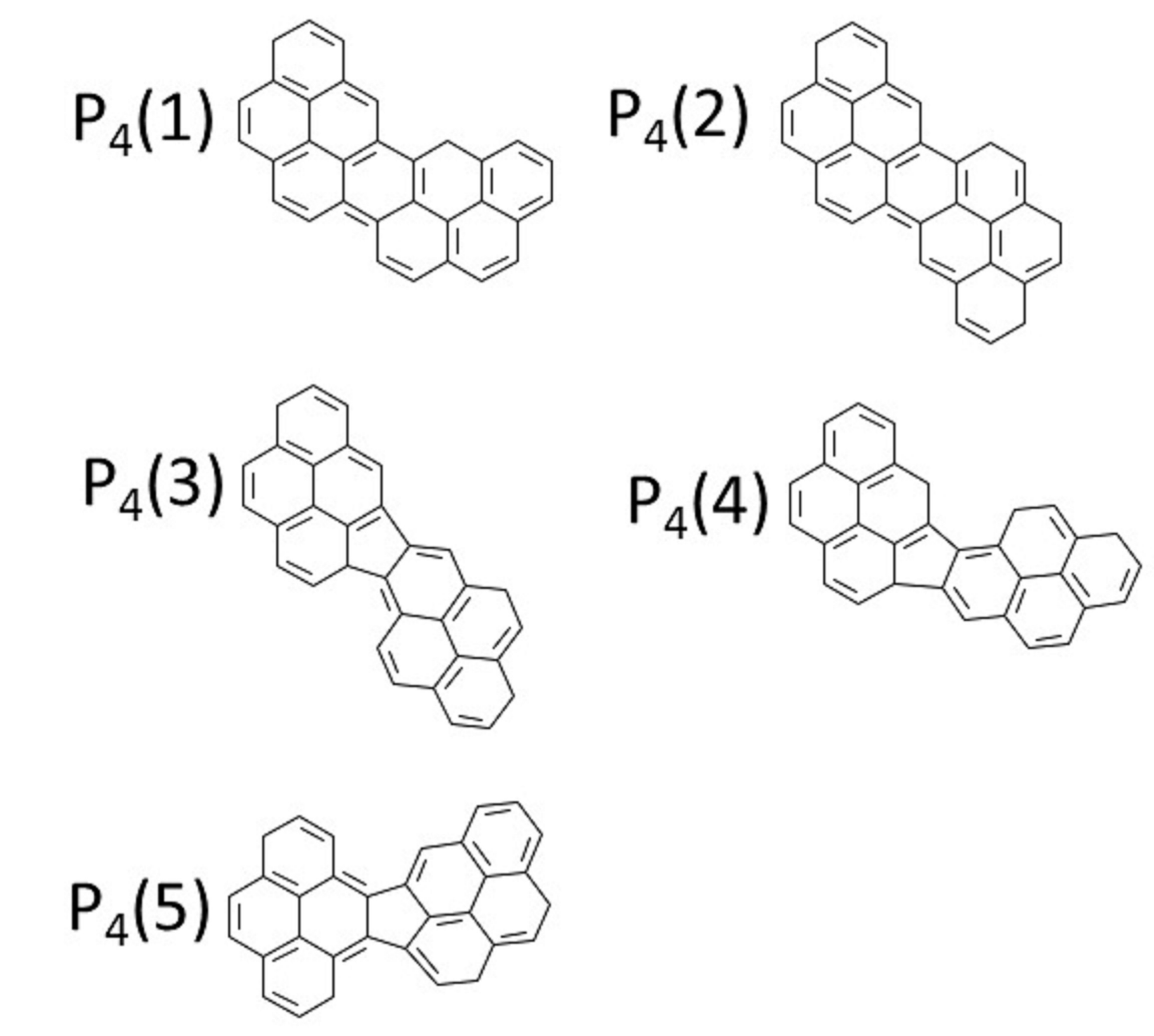}
 \caption{Possible isomers, C$_{32}$H$_{16}$$^+$, P$_4$(1-5), with planar structure that may have formed in our experiments. 
 }
 \label{fig4}
\end{figure}

\section{Experimental Methods}
\label{sec:exp}

The experiments are performed using the instrument for Photo-dynamics of PAHs (i-POP) \citep{zhen2014a}. We have slightly modified the set up to facilitate cluster formation. In short, neutral PAHs are transferred from a powder into the gas phase through heating in an oven (around 305 K).  An electron gun is applied to ionize the molecules. As a byproduct, some of the highly excited PAH may also lose an H-atom. A steel mesh (hole diameter $\sim$ 0.1 mm) is mounted on top of the oven to increase the local density of pyrene molecules, ions, and fragments. The increased density facilitates cluster formation. Recent studies on neutral, van der Waals bonded, acetylene clusters reveal that upon ionization, covalently bonded species are formed \citep{sti17}. Supporting quantum chemical calculations show that, in this process, solvent acetylene molecules assist in the barrier crossing dynamics on the potential energy surface and provide evaporative cooling that dissipates excess internal energy and stabilizes products. A similar process may be at work in our experimental set-up, where neutral pyrene clusters are formed in the high density regime above the mesh. These are then ionized by the electron gun and the excess energy facilitates ion-molecule reactions between cluster-monomers. The cluster is then cooled by evaporation of remaining, van der Waals bonded cluster members, leaving the covalently bonded product behind. We expect that clusters that do not undergo ion-molecule reaction with covalent bond formation will fragment completely in their constituent monomers as, for small clusters, the van der Waals bonding between pyrene molecules is only $\sim$ 0.4 eV per molecule \citep{rapacioli05b}. The ionized species are transported into the ion trap. After a short time delay (typically $\sim$ 0.2s), the SWIFT technique is applied to isolate species within a given mass/charge (m/z) range \citep{dor96}. The working pressures in QIT and TOF chamber are $\sim$ 8.0 x 10$^{-7}$ and $\sim$ 3.0 x 10$^{-9}$ mbar, respectively. The trapped ions are then irradiated by many (typically $\sim$ 5) pulses from a pulsed dye laser system. The content of the ion trap is then released and analyzed by a reflection TOF mass spectrometer. We have selected pyrene (C$_{16}$H$_{10}$, m/z=202;  Aldrich, 98 \%), for this study given its potential astrophysical interest \citep{tie05}, and the presence of armchair and zigzag edges which may facilitate the formation of new molecules \citep{poa07, kos08}. Typical experimental results are summarized in Fig. 1 and 2.

\section{Experimental results and discussion}
\label{sec:results}
A typical mass spectrum of pyrene dimer complex cations, after SWIFT isolation but before laser irradiation, is shown in Fig. 1(A). A series of the pyrene dimer complex cation is produced in our experimental condition. The inset provides more detail on the isotopologues and fragments. Given the high abundances of $^{13}$C containing pyrene dimer isotopologues (expected abundance ratio are $^{13}$C$_{0}$$^{12}$C$_{32}$:$^{13}$C$_{1}$$^{12}$C$_{31}$:$^{13}$C$_{2}$$^{12}$C$_{30}$$=$70.2:25.0:4.3), the mass spectra should be interpreted with care. Integration and calculations are necessary to obtain the abundances of pure $^{12}$C containing species. Following the method in \citep{zhen2014a}, we have used the natural isotope $^{12}$C /$^{13}$C abundance ratio (89) to deduce the abundance of the pure-$^{12}$C component of each species. These are then normalized to the total ion number. This results in: C$_{32}$H$_{15}$$^+$ (0.2 \%), C$_{32}$H$_{16}$$^+$ (6.6 \%), C$_{32}$H$_{17}$$^+$ (17.5 \%), C$_{32}$H$_{18}$$^+$ (22.8 \%), C$_{32}$H$_{19}$$^+$ (50.7 \%) and C$_{32}$H$_{20}$$^+$ (2.2 \%). These values show that under our experimental conditions, the (van der Waals bonded) pyrene dimer cluster cation (2.2 \%) is barely formed. The dehydrogenated pyrene cluster cations, i.e., partially H-stripped dimer ion species (pyrene dimer complex cation, e.g., C$_{32}$H$_{19}$$^+$, m/z=403, 50.7 \%) is the most abundant species. 

Fig. 1(B) shows the resulting mass spectrum of trapped pyrene dimer complex cations upon 595 nm irradiation at 0.2$-$1.0 mJ laser energies (irradiation times amounting to 0.5 s; i.e., typically $\sim$ 5 pulses). The mass spectra reveal a variety of fragment ions. The terminal photo-fragmentation pattern clearly depends directly on the incident radiation flux. The observed trend shows that, at low energy (e.g., 0.2 mJ), the pyrene dimer complex cations continue to dehydrogenate. With increasing laser energy, the peaks due to multiple fragmentation steps become more prominent and some fragmentation channels (carbon units loss) become accessible. 

Further details on the photo-dissociation behavior of the pyrene dimer complex cations are presented in Fig. 2, before laser irradiation and with 0.2 (lower laser energy) and 1.0 mJ (higher laser energy) irradiation. The dehydrogenation and carbon-unit loss behavior are illustrated in detail in Fig. 2(A) and Fig. 2(B), respectively. As before, we have corrected for the presence of the $^{13}$C isotopes in our analysis. In the range of m/z=401$-$406, we find that the intensity of dimers with odd number of H atoms is higher than those with even H numbers; e.g., the intensity of C$_{32}$H$_{19}$$^+$, m/z=403, and C$_{32}$H$_{17}$$^+$, m/z=401 are stronger than their neighbor, C$_{32}$H$_{18}$$^+$, m/z=402. On the other hand, in the range of m/z=390$-$400, we observe that the intensity of dimers with even H atoms is higher than those with odd H's, e.g., the intensity of C$_{32}$H$_{14}$$^+$, m/z=398 is stronger than its neighbors (C$_{32}$H$_{15}$$^+$, m/z=399 and C$_{32}$H$_{13}$$^+$, m/z=397). The different photochemical dehydrogenation behavior in these two regimes suggests that, after dehydrogenation, the pyrene dimer complex cations (e.g., [C$_{16}$H$_{10}$$-$C$_{16}$H$_{9}$]$^+$ or C$_{32}$H$_{19}$$^+$, m/z=403) convert to fully aromatic species (e.g., C$_{32}$H$_{16}$$^+$, m/z=400), i.e., PAHs. We will compare these experimental results with theoretical calculation later.

After these first H-loss steps and the (putative) conversion to aromatic species, the subsequent fragmentation pattern is that of typical PAHs as exemplified in the odd-even H-loss pattern \citep{zhen2014b,cas18}. In addition, carbon (CH or C$_2$H$_2$) loss channels are also identified in the mass spectra. Specifically, the fragments C$_{31}$H$_{15}$$^+$, m/z=387 and C$_{30}$H$_{14}$$^+$, m/z=374 are observed in Fig. 2 (A) already at 0.2 mJ laser energy. With the increasing laser energy (Fig. 2(A), 1.0 mJ), further dehydrogenation products with carbon loss (e.g., C$_{30}$H$_{8}$$^+$, m/z=368) are observed. In Fig. 2(B), a series of sequential CH or C$_2$H$_2$ loss and dehydrogenation photo-products are observed in the range of m/z=216-432 at 1.0 mJ irradiation. We observe masses corresponding to C$_{32-m}$H$_n$$^+$ with $m=[0-10]$ and $n=[0-8]$. We emphasize that these ions are not fully dehydrogenated and no pure carbon clusters are produced. The intensity of C$_{32-m}$H$_n$$^+$  with $m=2, 4, 6, 8, 10$ is stronger than its neighbors C$_{32-m}$H$_n$$^+$ with $m=1, 3, 5, 7, 9$, which suggests that the C$_2$H$_2$ loss channel is more favorable then the CH$+$CH loss channel or, alternatively, the second CH is more easily lost than the first CH \citep{eke98}. This carbon unit loss fragmentation behavior is similar to PAHs with the same size such as diindenoperylene, C$_{32}$H$_{16}$ \citep{cas18}. 

\section{Theoretical calculation results and discussion}
\label{sec:theoretical}
The theoretical calculations are carried out using density functional theory (DFT) with the hybrid density functional B3LYP \citep{bec92, lee88} as implemented in the Gaussian 16 program \citep{fri16}. All structures are optimized using the 6-311++G(2d,p) basis set. The vibrational frequencies are calculated for the optimized geometries to verify that these correspond to minima on the potential energy surface (PES). The dispersion-corrected B3LYP-D3 \citep{gri11} is considered to account the intermolecular forces in pyrene clusters. For these specific dimer masses, only a regular dehydrogenation scheme applies (i.e., not the deviating aliphatic one). In Fig. 3, we present the optimized structure of pyrene dimer complex cations that are likely formed in i-PoP, before and after laser irradiation and we suggest an evolution from the structures P$_{0}$ to P$_1$ (before irradiation) and P$_{1}$ to P$_4$ (after irradiation). 

First of all, it is necessary to determine the molecular structure of the pyrene dimer complex cation, C$_{32}$H$_{19}$$^+$, m/z=403, which we choose as starting point and assume that the smaller ones are dissociation products. On the basis of the theoretical calculations, we consider that the structure of C$_{32}$H$_{19}$$^+$ is that of two mono-pyrenes connected by a C$-$C single bond (as  [C$_{16}$H$_{10}$$-$C$_{16}$H$_{9}$]$^+$), where one carbon is in a sp$^2$ hybrid orbital, and the other carbon is in a sp$^3$ hybrid orbital with one C-H bond. The optimized carbon skeleton of these two pyrenes are not in the same plane, but form a three-dimensional structure, as shown in Fig. 3 (P$_1$). This radical cation is not completely aromatic, in agreement with our interpretation of the experimental photo-dehydrogenation results. 

Since the density of neutral and ionic pyrene is higher in the ionization zone of i-PoP, the possible formation pathway of C$_{32}$H$_{19}$$^+$ could be the interaction of neutral pyrene with a dehydrogenated pyrene cation. In the C$_{32}$H$_{19}$$^+$ formation process, there are three isomers of the dehydrogenated pyrene cation (C$_{16}$H$_{9}$$^+$; and three relative orientations of the neutral pyrene (C$_{16}$H$_{10}$) with respect to the dehydrogenated site of the pyrene cation available. As a result, there might be nine different possible isomer structures for the single C$-$C carbon band formation, linking the neutral and the cation in the resulting dimer, C$_{32}$H$_{19}$$^+$. One typical isomer structure is illustrated in fig. 3. The lowest energy formation pathway, P$_{0}$ $\longrightarrow$ P$_1$, is exothermic ($-2.95$ eV).

In the subsequent dissociation pathway initiated by the laser, P$_1$ to P$_4$ (Fig. 3), there is an activation energy of 1.81 eV for H loss from the aliphatic carbon (from P$_1$ to P$_2$). In contrast, H loss from the pure aromatic carbon rings takes around 5.0 eV, \citep{chen2015}. The next steps have an activation energy of 2.57 eV for H loss and formation a CC bond (from P$_2$ to P$_3$) and an activation energy of 2.34 eV for H loss from the aliphatic carbon (from P$_3$ to P$_4$), respectively. 

As showed in fig. 3, once the pyrene dimer complex cation (e.g., P$_0$) formed, with visible laser irradiation, the photo-process from P$_1$ to P$_4$ goes very efficiency, since the energy barrier for each step is lower (1.81, 2.57 and 2.34 eV) compare to the H-atom loss from pure aromatic carbon rings (5.0 eV). Mainly reason is the whole photo-process or the dehydrogenation process as an aromatization process, which will favor the conversion efficiency from P$_1$ to P$_4$, and favor the large PAHs formation through this routine. More theoretical studies regarding the possible transition states and dynamical processes will be discussed in a subsequent paper (Chen et al. 2018, in prep.). 

In addition, here, we stress that given the different isomeric starting structures of the C$_{32}$H$_{19}$$^+$ dimer, there might be five planar isomer structures of C$_{32}$H$_{16}$$^+$ (fig. 4), reflecting the molecular structure of pyrene with its two type of edge structures, zigzag and armchair. Thus, P$_4$(1) and P$_4$(2) is formed by combining one pyrene zigzag edge with another pyrene zigzag edge, while P$_4$(3), P$_4$(4) and P$_4$(5) are formed by combining one pyrene zigzag edge with one pyrene armchair edge. This illustrates the important role of the PAH edge structure in the bottom-up evolution routes towards large PAHs. 

\section{Astronomical implications}
\label{sec:discussion}

In summary, the experiments reveal a bottom-up formation process of large PAHs from clustered small PAHs (e.g., pyrene) and provide new insights into the evolution of PAHs in strong radiation fields. Consider the prototypical PDR in the reflection nebula, NGC~7023, observations show that some 3 \% of the elemental carbon is locked up in PAH clusters deep in PDRs, such as NGC 7023 \citep{rapacioli05a, tie08}. The young Herbig Be star, HD~200775, has created a cavity in the surrounding cloud that is broken open to the surrounding ISM \citep{fue98}. The strongly illuminated molecular cloud to the North shows the characteristics of a bright PDR with prominent emission in the ro-vibrational and pure rotational H$_2$ lines, the atomic fine-structure lines of [CII] and [OI], and typical molecular tracers of PDRs \citep{joblin10,martini97,fue98}. The mid-IR emission from this region is dominated by the well-known PAH emission bands. Analysis of these spectra reveals three distinct emission components and the characteristics of these components have been attributed to PAH cations, PAH neutrals, and PAH clusters, respectively \citep{berne07}. These components show different spatial distributions in this source and present a prime example of PAH evolution. This data has been interpreted that, as the gas flows through the PDR and into the cavity, it is exposed to increasingly stronger radiation fields. As a result, first, PAH clusters will fall apart. Subsequently, PAHs are converted into their most stable forms (GrandPAHs), and then destroyed or converted into C$_{60}$ and other cages \citep{berne12, zhen2014b,and2015}. The experiments presented here modify this scenario by introducing the conversion of van der Waals bonded PAH clusters into larger PAHs as a process counteracting the general decrease in PAH size in this flow.

Studies of the typical size of the emitting PAHs have revealed a complex PAH size evolution with distance from the illuminating star in the PDR of NGC 7023 \citep{croiset16}. This size evolution is accompanied by an evolution in the aliphatic-to-aromatic ratio as evidenced in the 3.4/3.3 $\mu$m ratio. This spectral evolution has been interpreted as evidence for photo-processing of PAH clusters leading to the formation of (large) PAHs with aliphatic side groups deep in the PDR and loss of these aliphatic groups closer to the surface of the PDR \citep{pilleri}. The large PAH formation mechanism obtained in here maybe stand for the point of PAH clusters evolution. Further observational and experimental studies are required to address these links. 

\section{Conclusions}
\label{sec:concl}
We have presented the first experimental results on the conversion of small PAH clusters (e.g., pyrene clusters) to large PAHs process in the gas phase under the influence of laser irradiation.  We summarize our results as follows: (1) Our experiments reveal a general, photo-driven process that converts PAH-clusters into large PAHs. This photo-processing provides a reasonable explanation for the formation of large PAHs in space; (2) The formation of large PAHs from PAH-related clusters in ISM is very efficient, as it boils down to an aromatization process; (3) Quantum chemical calculations demonstrate that PAH edges (e.g., armchair and zigzag edges) play an active role in the formation process and control the specific PAHs formed. Consequently, the structure of large PAHs initially formed by this process is diverse. Subsequent photo-processing may further weed down those PAHs to their most stable forms. Further experimental studies are warranted, as the result presented in here only can apply to small PAH clusters, the photo-dissociation process of larger PAH clusters is still valid.

\acknowledgments
This work is supported by the Fundamental Research Funds for the Central Universities and from the National Science Foundation of China (NSFC, Grant No. 11743004). TC acknowledge the Swedish Research Council (Contract No. 2015-06501), calculations are carried out on the supercomputers supported by the Swedish National Infrastructure for Computing (SNIC). AT acknowledge the European Union (EU) and Horizon 2020 funding awarded under the Marie Skłodowska Curie action to the EUROPAH consortium, Grant No. 722346. Studies of interstellar PAHs at Leiden Observatory are supported through a Spinoza award.

\end{document}